# Ferroelectric Nematic Droplets in their Isotropic Melt

*Kelum Perera[1,2], Rony Saha[1,2], Pawan Nepal[4], Rohan Dharmarathna[2], Md Sakhawat Hossain[2,3], Md Mostafa[2,3], Alex Adaka[2,3], Ronan Waroquet[2], Robert J. Twieg[4] and Antal Jákli[1,2,3]*

[1]Department of Physics, Kent State University, Kent OH, 44242, USA

[2]Advanced Materials and Liquid Crystal Institute, Kent State University, Kent OH, 44242, USA

[3]Materials Science Graduate Program, Kent State University, Kent OH, 44242, USA

[4]Department of Chemistry and Biochemistry, Kent State University, Kent, OH 44242, USA

## Abstract

The isotropic to ferroelectric nematic liquid transition had been theoretically studied over one hundred years ago, but its experimental studies are rare. Here we present polarizing optical microscopy studies and theoretical considerations of ferroelectric nematic liquid crystal droplets coexisting with the isotropic melt. We find that the droplets have flat pancake-like shapes that are thinner than the sample thickness as long as there is a room to increase the lateral droplet size. In the center of the droplets a wing shaped defect with low birefringence is present that moves perpendicular to a weak in-plane electric field, and then extends and splits in two at higher fields. Parallel to the defect motion and extension, the entire droplet drifts along the electric field with speed that is independent of the size of the droplet and is proportional to the amplitude of the electric field. After the field is increased above $1V/mm$ the entire droplet gets deformed and oscillates with the field. These observations led us to determine the polarization field and revealed the presence of a pair of positive and negative bound electric charge due to divergences of polarization around the defect volume.

## 1. Introduction

The isotropic-nematic phase transition has intrigued physicists in the early 20$^{th}$ century soon after the discovery of liquid crystals at the end of the 19$^{th}$ Century[1,2]. Notably, the first attempt to explain this transition was given by Max Born, who later received a Nobel prize for his work in quantum mechanics. In 1916 Born proposed that nematic liquid crystals are ferroelectric in which the strong dipole–dipole interactions among molecules stabilize the orientational order against thermal fluctuations[3]. The condition for the molecular dipole moment $\mu$ can be estimated[4] to be $\frac{\mu^2}{\varepsilon_0 \varepsilon V_m} > k_B T$, where $V_m$ is the molecular volume, $k_B$ is Boltzmann constant and $T$ is the temperature. At room temperature with $V_m \sim 1nm^3$ and $\varepsilon \sim 10$, this estimate gives that the molecular dipoles must be larger than $\mu \sim 6\ Debye$. Experimentally, ferroelectric nematic fluids were found about a century after Born's paper by Nishikawa et al.[5] and Mandle et al.[6,7] in materials containing rod-shaped molecules with dipole moments of about $10\ Debye$. In these materials (called DIO and RM734, respectively) the ferroelectric phase did not form directly below the isotropic phase, but below an intervening non-ferroelectric nematic phase. These publications inspired many groups worldwide to study the nature of ferroelectric nematic ($N_F$) phase of liquid crystals as summarized in a perspectives article by Sebastián et al.[8] The $N_F$ phase with polar order[9,10] has unique viscoelasticity[11], topology[12,13], and electrooptical properties[14–16] that may play a key role in many future technological advances like data storage, sensors, mechanical actuators, displays with sub-millisecond switching, and other optoelectronic applications[15–17].

In the past five years many $N_F$ materials have been synthesized but only a few were found to exhibit a direct isotropic to ferronematic transition[18–20]. In these materials the $N_F$ phase forms via nucleation where polarized optical microscopy (POM) images revealed circular-shaped domains that were assumed to be spherical[15]. On cooling in the $Iso + N_F$ two phase range, the nuclei gradually grow, merge, and eventually fill the whole area in a mosaic-like pattern in the $N_F$ phase[8]. Second Harmonic Generation (SHG) and interferometric microscopy studies showed[11] that in the circular domains the nematic director forms a +1 defect with a concentric pattern, for which tangential clockwise and anticlockwise directions of polarization can be found with equal probability[11,15]. Such observations are consistent with polarimetric observations by Máthé et al. on $N_F$ sessile droplets in air[12].

In this paper we present detailed studies of the $N_F$ nuclei in various electric fields on a $N_F$ material with a direct $Iso - N_F$ transition. We found that the initial spherical nuclei grow radially sideway to form pancake shapes that can float in the isotropic fluid. In in-plane electric fields the central defect lines move perpendicular to the electric field, while the entire domains move along the applied field. Based on these observations we determine the polarization field and show the presence of positive and negative bound electric charges due to divergences of polarization around the defect area.

## 2. Material and methods

The synthesis and chemical characterization of RT11064, Benzoic acid, 2,4-dimethoxy-, 3-fluoro-4-[(3-fluoro-4-nitrophenoxy)carbonyl]phenyl ester, is provided in the Supporting Information (SI) section. The molecular structure of RT11064 and the DSC curves at the third heating-cooling cycle are shown in Figure 1. RT11064 is identical to compound 3 of Reference[21].

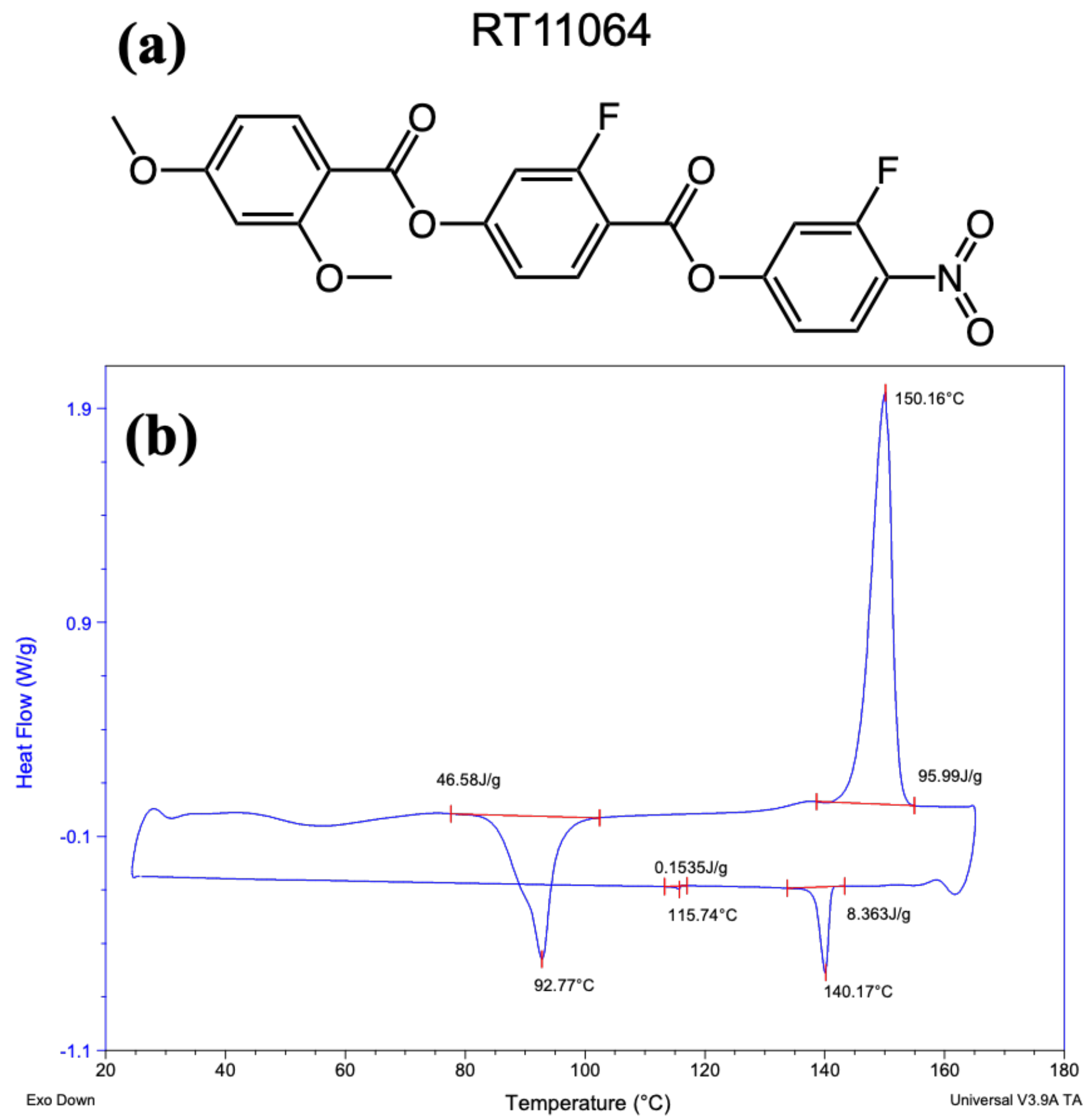

*Figure 1: Molecular structure of RT11064 (a) and the DSC curves at third heating-cooling cycle (heating and cooling rate = 5 ˚C/min. Sample size = 5.567 mg) indicating a phase sequence: Iso – 140°C – (Iso+$N_F$) – 115°C $N_F$ – Cr (93°C).*

For polarized Optical Microscopy (POM) $5 - 10\ \mu m$ thick samples were placed in an Instec HS200 heat stage and viewed through crossed polarizers with an Olympus BX60 microscope. Electro-optical and ferroelectric polarization measurements were carried out in cells with rubbed

PI-2555 polyamide coating and in one plate with two 2 $mm$ wide conducting ITO strips separated by 1 $mm$ distance. The cells were assembled using NOA68 glue with appropriate spacers to achieve uniform thickness. An HP 33120A function generator amplified by FLC F10A 20X amplifier from Instec were used to apply AC voltages.

For the polarization measurements a 10 $\mu m$ thick sample was studied under 200 $Hz$ triangular electric fields. Since the gap between the electrodes is much larger than the film thickness, we assume that the net polarization charge $Q = P_o A$ appearing on the in-plane electrodes is independent of the size of the electrodes[9] and is determined only by the length of the electrodes multiplied by the film thickness ($A = 1cm \times 10\mu m = 10^{-7}\ m^2$) normal to the polarization vector.

## *3.* Experimental Results

Representative polarized optical microscopy (POM) textures of a 6 $\mu m$ cell is shown Figure 2a on cooling from the isotropic phase with rate of 1 $°C/min$. The $N_F$ phase nucleates from the isotropic phase as circular-shaped domains. The radii of the very first domains are less than the film thickness, but then they grow laterally with radii exceeding the film thickness until they merge as shown in Figure 2a. The two-phase range was found to depend on the film thickness and cooling rate: the thinner the film and slower the cooling, the wider is the range. The birefringence color is almost temperature independent in the two-phase region, then abruptly increases once the domains merge to form the pure $N_F$ phase where it is only slightly increasing on cooling. In the two-phase range (top part of Figure 2a) the optical path difference can be estimated from the Michel-Levy chart (see top part of Figure 2b) to be about 1050 $nm$, whereas in the pure $N_F$ phase it is about 1250 $nm$.

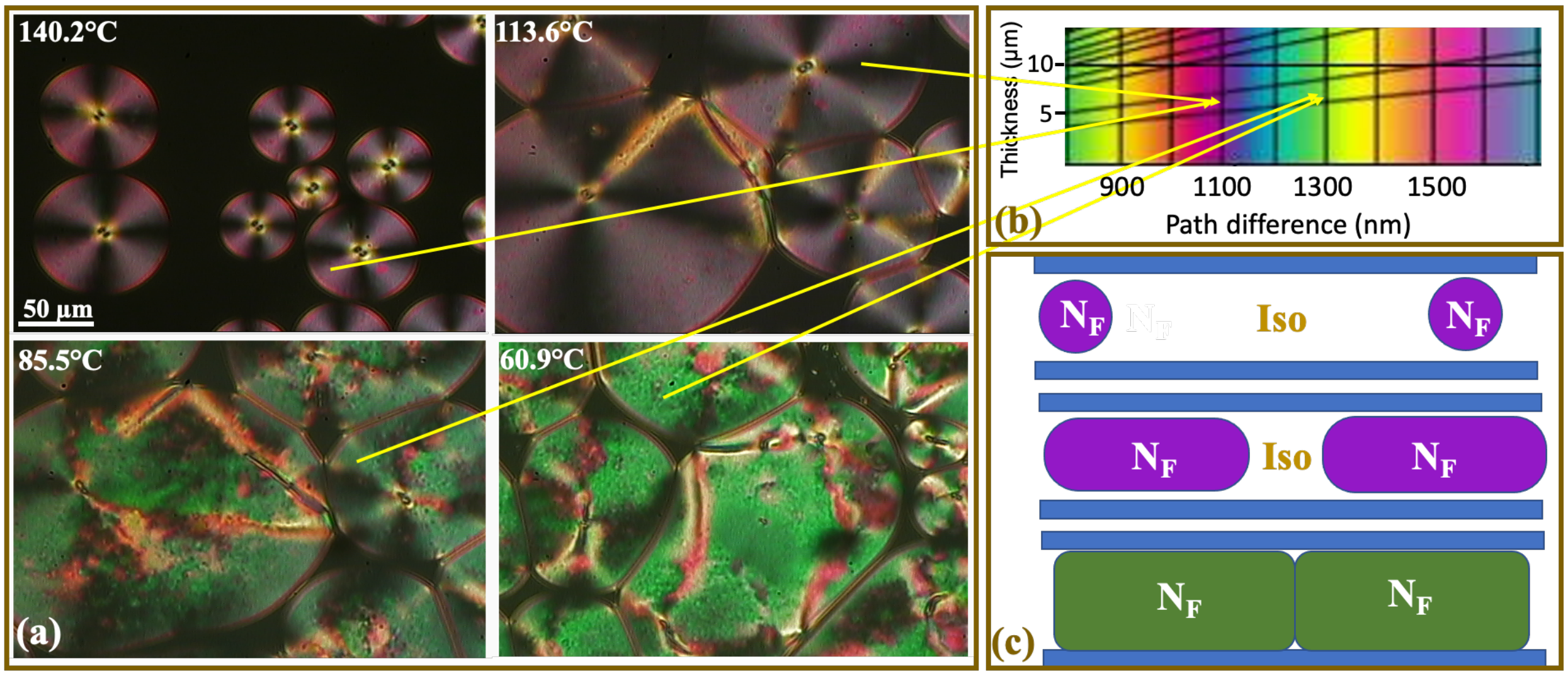

*Figure 2: (a): Polarized Optical Microscopy (POM) images at 4 different temperatures of a 6 µm thick RT11064 film on cooling with 1 °C/min rate from the isotropic phase. (b): A portion of the Michel-Levi chart with the birefringence colors corresponding to the POM colors at different temperatures. (c): Illustration of the approximate cross section of the phase structures in the Iso-$N_F$ range and in the $N_F$ range, based on the POM images in (a).*

Assuming planar alignment, we estimate the birefringence in the $N_F$ phase to be about $1.25\ \mu m/6\ \mu m \sim 0.21$. With a slightly smaller birefringence we estimate the thickness of the $N_F$ nuclei to be $1.05/0.20 \sim 5.2\ \mu m$, i.e., smaller than of the film thickness. Illustration of the side-view of the $N_F$ domains right after the first spherical nuclei appear, of the pancake shaped nuclei in the middle of the $Iso + N_F$ two phase range, and the fused $N_F$ domains in the pure $N_F$ range in the top, middle and bottom sections of Figure 2c, respectively.

The temperature dependence of the ferroelectric polarization of RT11064 with representative time dependences of the polarization current in the inset is shown in Figure 3. The 118°C-140°C temperature range corresponds to the two-phase range where the $N_F$ volume of the nuclei increase. In the pure $N_F$ range the polarization linearly increases from about $6.5\ \mu C/cm^2$ to $7.8\ \mu C/cm^2$. The dotted line indicates the temperature dependence of the ferroelectric polarization of the individual $N_F$ nuclei in the two-phase range.

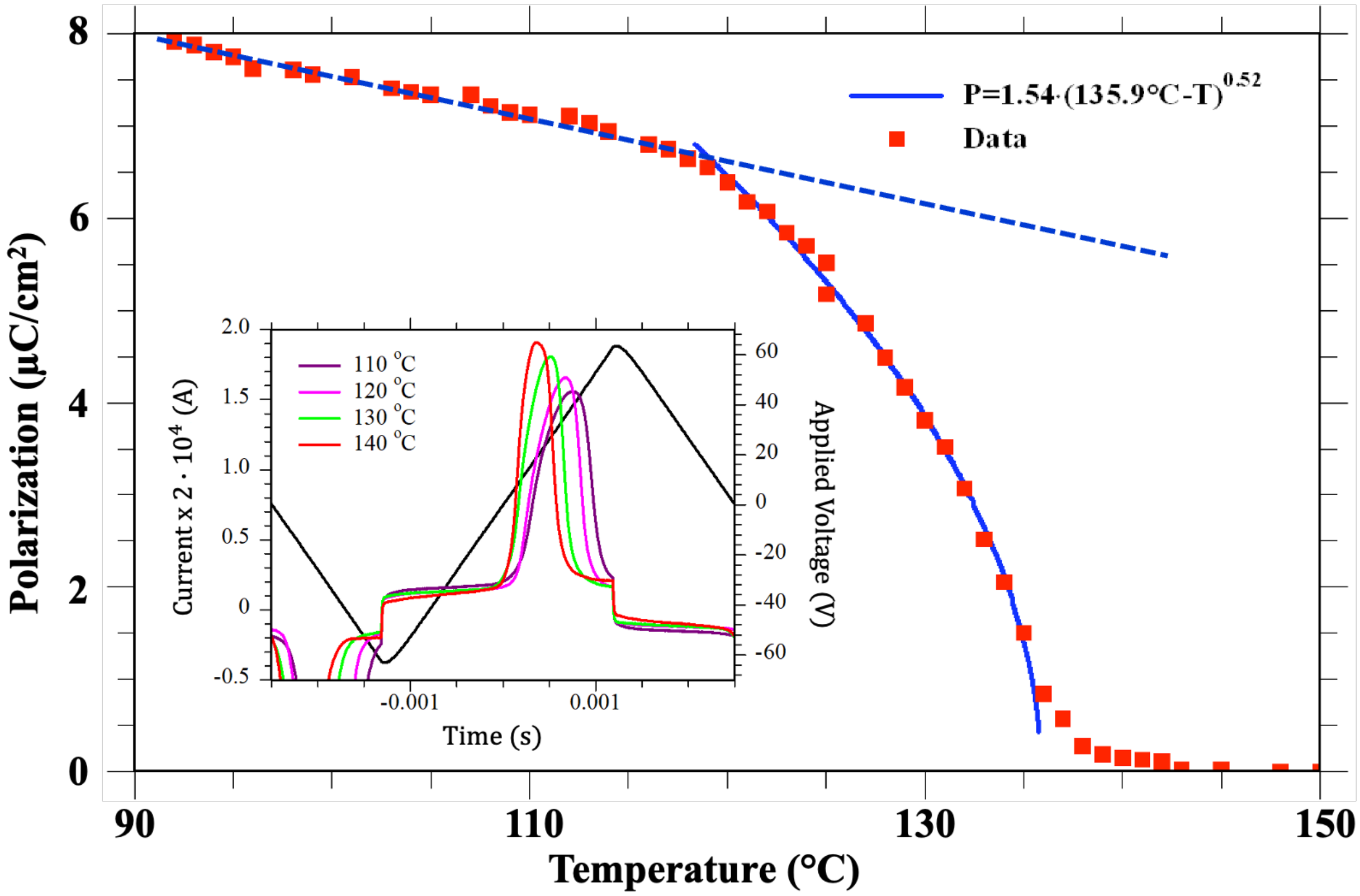

*Figure 3: Temperature dependence of the spontaneous polarization of RT11064. Inset: Time dependence of the applied voltage (Right axis) and measured voltage dropped on* $20k\Omega$ *resistance (left axis) at selected temperatures*

POM images of representative $N_F$ nuclei in various electric fields are shown in Figure 4. Figure 4 (a and b) compare the structures and positions of two nuclei at zero electric field and 3seconds after application of a $0.6\, mV/\mu m$ electric field in the $x$ direction. One can see two simultaneous motions. Initially, the central defect moves perpendicular to the electric field by a distance $\Delta y \approx 10\mu m$ for the top $D \approx 50\mu m$ diameter nucleus and by $\Delta y \approx 7\mu m$ for the bottom $D \approx 35\mu m$ diameter nucleus. At the same time the entire nuclei move along the electric field by approximately the same $\Delta x \approx 16\mu m$ distance. While all droplets moved in the same direction, the defect in the center of each domain moved perpendicular to the electric field either left or right as shown in Figure 4b. The velocity of the droplets is found to be independent of the rubbing direction, which indicates that these droplets do not touch the substrates. Under weak DC voltages all droplets were found to move the in the electric field toward the negative electrodes.

Figure 4c shows a $D \approx 90\mu m$ diameter circular shaped nucleus at zero, $0.4\, mV/\mu m$ and $0.6\, mV/\mu m$ electric fields from top to bottom in $d = 5.5\ \mu m$ thick film between crossed polarizers along $x$ and $y$ directions. We see the nucleus has second order magenta color outside the defect area, black in the defect and first order whitish to second order green colors around the defect. Comparing with the Michel-Levy chart, we see that the magenta color corresponds to an optical path difference of $\Gamma = \Delta n \cdot d \approx 1100\ nm$. From the $d = 5.5\ \mu m$ film thickness this provides that the birefringence of the $N_F$ nucleus is $\Delta n \approx 0.2$, which agrees with the estimate based on the POM images in Figure 2a. In the defect area black indicates zero birefringence, the whitish color corresponds to $\Delta n \sim \frac{0.3}{5.5} \sim 0.05$ birefringence and the green color corresponds to $\Delta n \sim \frac{0.7}{5.5} \sim 0.13$ birefringence.

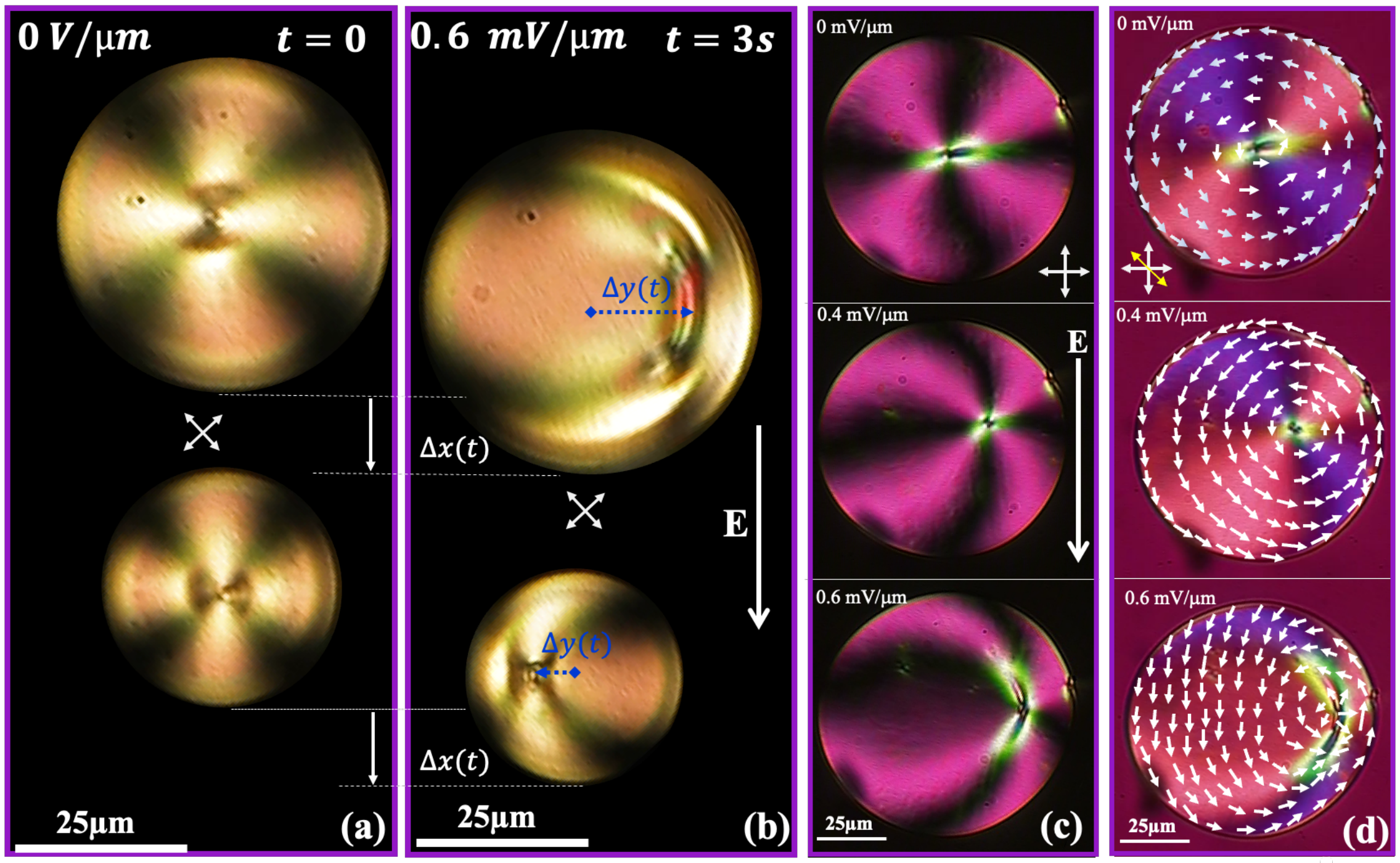

*Figure 4: POM images of $N_F$ nuclei formed at the $Iso - N_F$ transition without and with electric field applied in x direction. (a): Two different sized nuclei in a 10 μm thick film between crossed polarizers at $\pm 45°$ with respect to the $x - y$ axes at zero electric field. (b): The same nuclei as shown in (a), at $t = 3s$ after $E = 0.6 mV/\mu m$ DC electric field was applied in x direction. (c): One 90 μm diameter circular shaped nucleus at zero, 0.4 mV/μm and 0.6 mV/μm electric fields from top to bottom in 5.5 μm thick film between crossed polarizers in x and y directions. (d): The same nucleus as in (c) but now a $\lambda = 530\ \mu m$ waveplate inserted at 135° between the crossed polarizers. White arrows indicate the approximate ferroelectric polarization directions.*

Figure 4d shows the same nucleus at the same fields as in Figure 4c but now with a $\lambda = 530\ \mu m$ waveplate inserted at 135° between the crossed polarizers above the sample. When the optical axis of the LC is parallel to that of the waveplate, the resulting optical path difference far from the defects is $\Gamma_{\parallel} \approx (1100 + 530) nm = 1630\, nm$ corresponding to third-order light pink color. When the director of the LC is perpendicular to that of the waveplate, then $\Gamma_{\perp} \approx (1100 - 530)\, nm = 570\, nm$ corresponding to a second-order purple color. From these observations we can verify that, except for near the defect area, the director has tangential distribution in agreement with[11,15,12]. As we will discuss later, combining this information with the direction of the movement of the defect, we can also determine the direction of the polarization (see white arrows overlayed the POM images in Figure 4d), as it reasonable to assume that the field will deform the structure to provide a net polarization along the electric field. The change of colors around the defect area (whitish to green and from green to yellow) is also consistent with the previous conclusion that the alignment is homeotropic with zero birefringence in the core of the defect and the birefringence increasing slowly in one direction (along the wing), slowly (within 20 µm), and abruptly (within <3 µm) perpendicular to it. This trend indicates anisotropic tilt variation around the defect.

The voltage dependence of the amplitude of the side-wise motion of the defect for a 46µm diameter nucleus at 4°C below the appearance of the first nucleus measured in a 5µm cell is shown in Figure 5a at various frequencies between $500\, mHz$ and $1\, Hz$. One can see that up to $1V$ ($E < 1\, mV/\mu m$) the amplitude is proportional to the applied in-plane voltage. The top insets show that the motion follows well the sinusoidal variation of the applied voltage, while the lower inset illustrates the direction of the displacement with respect to the external applied field.

The motion of the droplets along the electric field is demonstrated in the Supporting Video 1. The electric field dependence of the velocity of a $D = 78\, \mu m$ diameter droplet under a square wave electric field is shown in Figure 5b for several frequencies. The inset at the top right corner of the figure shows that for 100 mHz square wave voltage the velocity is constant while the voltage is constant. The main pane of the figure shows that the velocity is proportional to the applied electric field and decreases at increasing frequencies. The frequency dependence in $0 - 2.6\, Hz$ interval of $1\, mV/\mu m$ electric field is shown in the top-left inset with the fitting equation $v = 3 + 7.4 \cdot e^{-f/0.47}$. This indicates a low frequency contribution with amplitude of $7.4\, \mu m/s$ that decays exponentially to $1/e$ value at $0.47\, Hz$, and there is a higher frequency component of $v_h \approx 3 \mu m/s$.

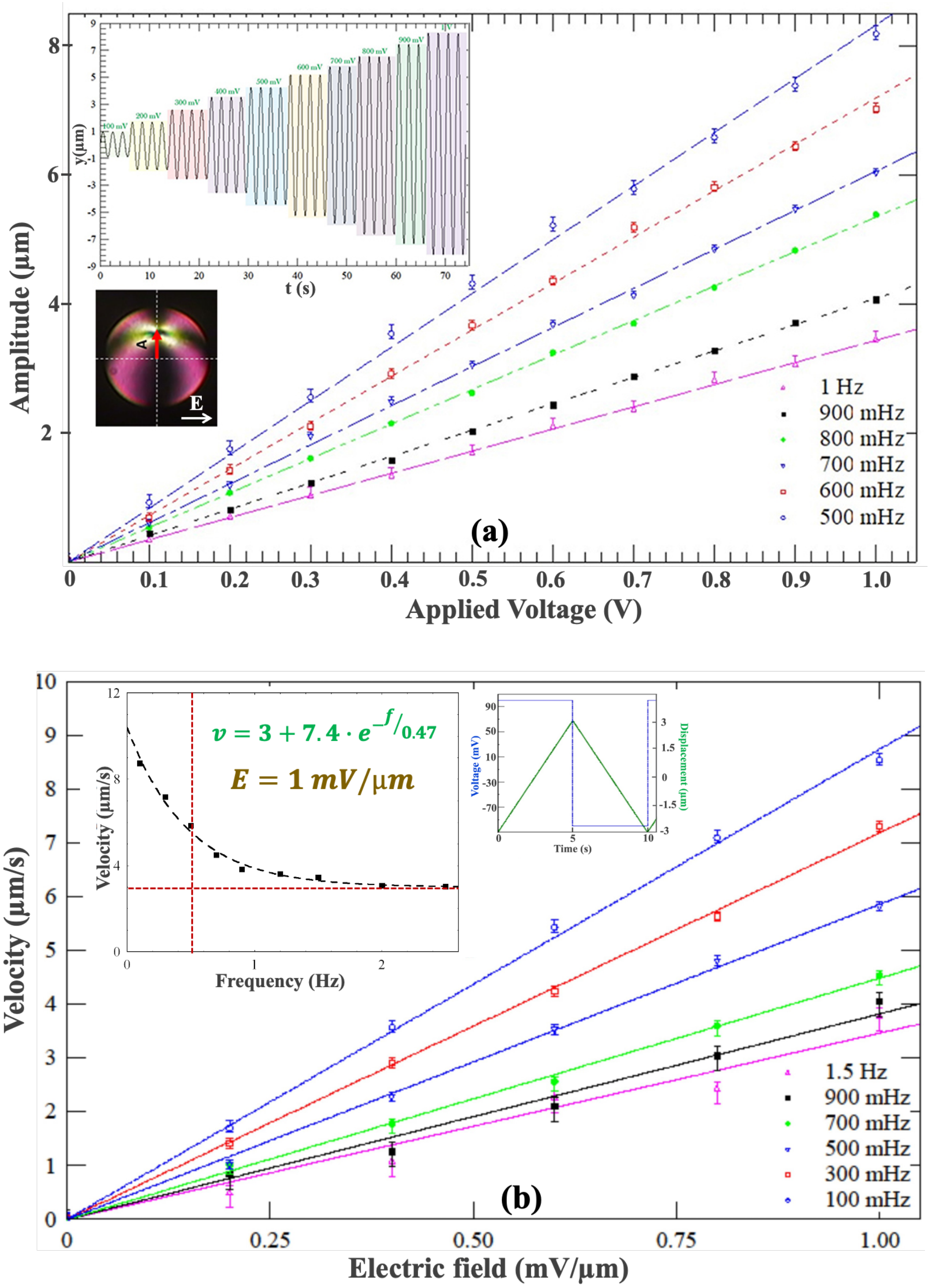

*Figure 5: Electric field dependence of the defect movement and of the velocity of the $N_F$ droplets. (a): Applied voltage dependence of the displacement of the central defect of a $D = 46$ μm diameter droplet in $\mathrm{d} = 5$ μm thick film at 4°C below the appearance of the first nucleus in the isotropic phase. Top-left inset shows the sinusoidal time dependence of the displacement at several frequencies. The lower inset illustrates the direction of the displacement with respect to the external applied field (b): Electric field dependence of the velocity of a $D = 78$ μm diameter $N_F$ nucleus for various square wave fields. Top right inset: Displacement under $100\ mHz$ square wave voltage. Top-left inset: Frequency dependence of the velocity in $0 - 2.6\ Hz$ interval at $1\ mV/\mu m$ electric field with the fitting equation.*

All the above measurements were done at low fields ($E < 1\,mV/\mu m$ or $f > 0.5\,Hz$). At increasing fields and/or decreasing frequencies, the defects not only moving inside the droplets, but they extend and split and simultaneously even the shapes of the droplets change.

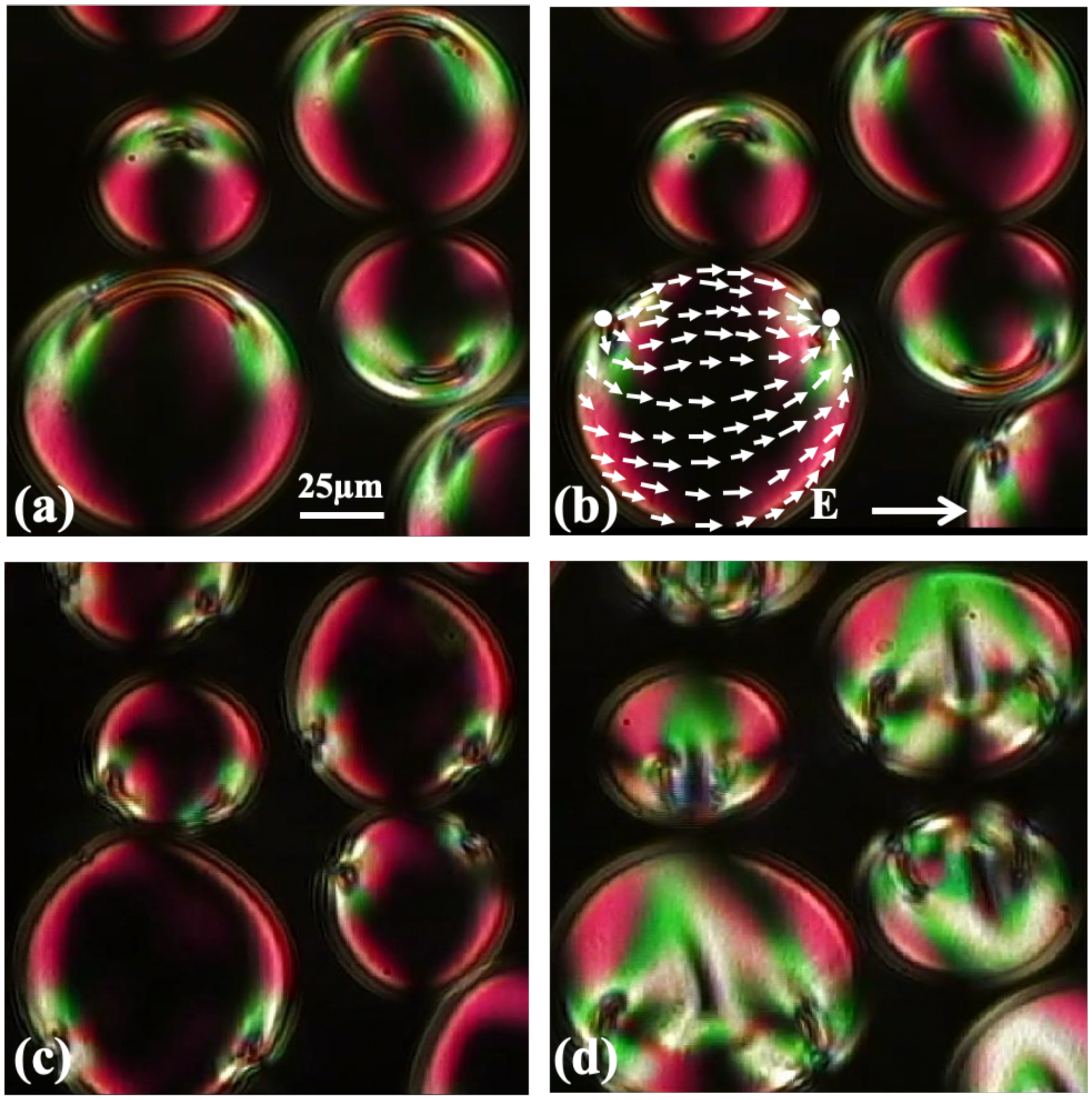

*Figure 6: Illustration of the behavior of the pancake-like droplets in high electric fields applied horizontally. (a and b): f=* $0.1\,Hz$ $E_m = 0.5\,mV/\mu m$ *(a) snapshot when* $E = 0.4\,mV/mm$*; (b) snapshot at* $E = 0.5\,mV/mm$*. (c and d): f=* $0.1\,Hz$ $E_m = 1.5\,mV/\mu m$*; (c)* $E = 0.5\,mV/\mu m$*; (d)* $E = 1.5\,mV/\mu m$ *. Crossed polarizers are in horizontal and vertical directions.*

The extension and the split of the defects along the applied are shown in Figure 6(a and b) when $f = 0.1\,Hz$ field is applied with $E_m = 0.5\,mV/\mu m$ amplitude. Figure 6a shows the extension of the defect at the instant when $E \sim 0.4 \frac{mV}{mm}$ field. Figure 6b shows snapshot at $E = 0.5\,mV/mm$ after the defect has split to two and situated at the opposite sides of the droplets. The approximate polarization field is overlayed on one of the droplets. At increasing amplitudes at the same $f = 0.1\,Hz$ frequency the droplet shape changes as shown in Figure 6(c and d) and in

Supporting Video 2. During the vertical motion of the defect and its horizontal extension the droplets elongate along the motion of the defect and reaches its highest aspect ratio right after the defect splits as shown in Figure 6c at $= 0.5\ mV/\mu m$. At further increasing fields the droplets start extending along the applied field and reach their highest aspect ratio at the highest field as shown in Figure 6d at $= 1.5\ mV/\mu m$. As it is apparent from the pictures of the different droplets in Figure 6, the magnitude of the fields where the above changes depend also on the size of the droplets: the smaller are the droplets, the higher is the field required for the above changes. The threshold fields also increase with increasing frequencies (for example, the threshold for splitting increases to 7 $V$ at 0.7 $Hz$).

## 4. Discussion

The results we presented above reveal several notable properties of the ferroelectric nematic liquid crystal droplets that coexist with the isotropic liquid phase. (i) The droplets have a flat pancake-like shape that are thinner than of the sample thickness if there is a room to increase the droplets laterally. (ii) In the center of the droplets a defect with wing shape darker pattern appears instead of the four Maltese crosses that are seen far away the central defect. (iii) In the presence of in-plane electric fields the central defect moves perpendicular to the electric field, then it extends along the field and at high enough field it splits to two defects. (iv) While decreasing the field after strong and low frequency fields applied, the circular base extends along the field. (v) In parallel to the defect motion and extension, the entire droplet drifts along the electric field with size independent speed that is proportional to the in-plane electric field. In this section we attempt to explain all the above features starting with (ii) which is needed to explain (i) and all other features.

As in $N_F$ $\hat{n} \neq -\hat{n}$, one cannot have half integer defects, which are the most stable in conventional nematics. The central defect therefore must have a strength of an integer number. As previous results on $N_F$ droplets in an isotropic melt[11,15] and in air[12] showed the director and the polarization has tangential distribution far away from the central defect, it is reasonable to assume a simple $+1$ central defect line as proposed by Li et al[20]. Such a structure however would result in a simple 4 brush Maltese cross along the crossed polarizers. This is clearly not true in our case as shown in Figure 2a and Figure 4a, where two-fold wing-type structures are seen. We propose this structure can be explained by the director and polarization field we show in **Error! Reference source not found.**.

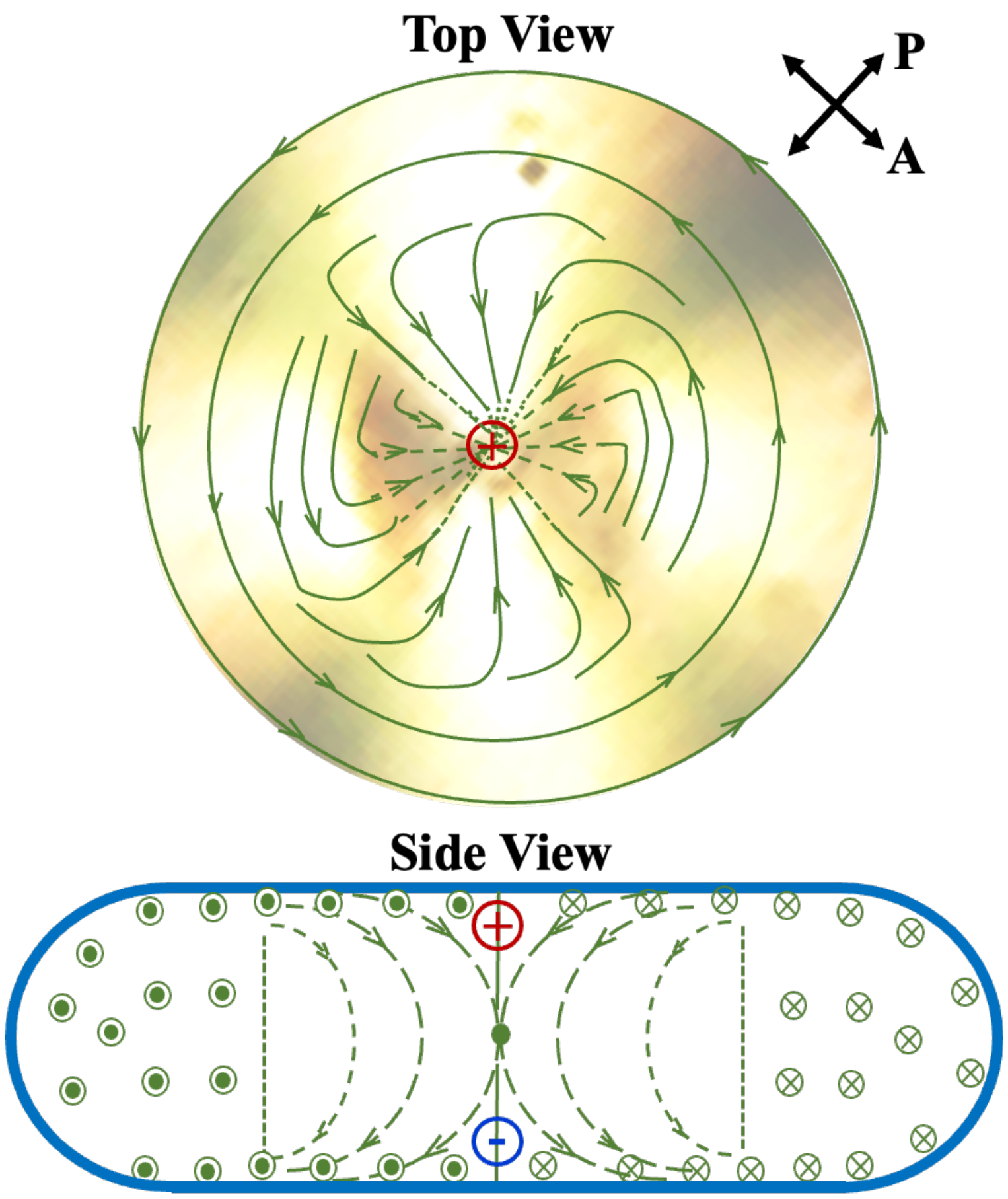

*Figure 7: Illustration of the director structure in the pancake-like $N_F$ domains. In the top view the director structure is drawn on the POM image of a part of an actual droplet shown in Figure 4a. Dotted lines indicate director tilting away from the plane of the drawing. Shorter dots mean larger tilt. A dot (cross) in a circle shows director perpendicular to the plane with polarization direction pointing toward (away) from the reader.*

The wing-type darker spots indicate an increasing deviation of the director field from the planar orientation toward the center. This leads to a $+1$ point defect in the middle of a $+1$ line defect running across the film as seen in the side view of the droplet. Due the divergence of the electric polarization, such a structure will lead to a bound charge density $\rho = -\vec{\nabla}\vec{P}$. Notably the divergence at the upper and lower parts are opposite (in the example shown in **Error! Reference source not found.**, inward in the upper part and outward in the bottom) leading to different signs of bound charges at the upper and lower parts (e.g., positive charge in the upper and negative charge in the lower part in **Error! Reference source not found.**). This funnel-like polarization divergence appears to have two-fold symmetry where the inward polarization divergence is concentrated along the East-West direction as in the top droplet of Figure 4a, or can be along South-North direction as in  as in the bottom droplet of Figure 4a. The situation corresponding to the former shown overlayed on the inner part of the POM of the top droplet of Figure 4a, is shown in the Top view of **Error! Reference source not found.**. Such opposite bond charges at the upper and lower parts of the droplet attract free ions with opposite signs of the isotropic liquid above and below the droplets. The screening of the bound charges therefore requires the presence of the isotropic liquid below and above the $N_F$ droplets, which might explain that an isotropic layer is required between the $N_F$ droplets and the substrates, leading to a lateral growth of the droplets.

The lateral motion of the central defect under weak electric fields inside the pancake shaped $N_F$ domains perpendicular to the applied voltage can be understood by the field induced polarization $P_E$ , as schematically illustrated in Figure 4d. For small displacement ($x \ll R$, where $R$ is the radius of the droplet) the induced net polarization can be approximated as $P_E \approx \frac{1}{wR^2\pi} \iint P_E (r, \varphi) r dr d\varphi \approx P_o \frac{x}{R\pi}$, where $w$ is the thickness of the disc. This will decrease the free energy of the ferroelectric domains by $\Delta W \approx -P_o x R w E$, where $E = \frac{V}{L}$ is the electric field due to the voltage $V$ applied across the in-plane electrodes with gap of $L$. The resulting force $F_E$ acting on the defect perpendicular in the field, is $F_E = -\frac{dW}{dx} = P_o R w E$. This force is balanced by the two viscous forces, the drag force $F_D$ acting on the defect line and $F_\gamma$ related to the rotation of the polarization in a distance $x$ and length $2R$. The drag force can be given as $F_D \sim \xi a \eta v$, where $a$ is the radius of the defect line, $\eta$ is the relevant flow viscosity of the ferroelectric fluid, and $\xi(a, w, R)$ is a geometric factor depending on the size and shape of the defect line and on the size of the disc shaped domain. (Note: According to Stokes low, for spherical defects in an isotropic and infinite fluid, $\xi = 6\pi$.) The rotational viscous force can be approximated as $F_\gamma \approx \frac{d}{dx}(2Rwx\gamma_1\dot{\varphi})$, where $\dot{\varphi} = \frac{d\varphi}{dt} = 2\pi/T$, where $T$ is the time with speed $v$ to translate by $R$ ($v = \frac{R}{T}, i.e., \dot{\varphi} = \frac{2\pi v}{R}$). Accordingly, we can write that $F_\gamma \approx 4\pi w \gamma_1 v$. In equilibrium $F_E = F_D + F_\gamma$, i.e., $P_o R w E \approx v(\xi a \eta + 4\pi w \gamma_1)$. This gives a speed of the sidewise motion of the defect line which is

$$v \approx \frac{P_o R w E}{\xi a \eta + 4\pi w \gamma_1}. \qquad (1)$$

Experimentally we indeed see that the speed is proportional to weak applied field and voltage and the maximum displacement relative to the radius of the disc shape domains $\frac{xf}{R} = v/R$ is constant as seen in Figure 4a. For a quantitative comparison with the experiment, we take from Figure 5a that at $f = 1\,Hz$ and $E = \frac{1V}{1\,mm} = 10^3 V/m$, $x \sim 4\mu m$. The polarization of the droplets can be extrapolated (see dashed blue line of Figure 3) to be $P_o \sim 5 \cdot 10^{-2} C/m^2$. Taking $w \sim 4\mu m$, $a \sim 10\mu m$, $R = 23\mu m$ and $\xi \sim 6\pi$, we get the right order of magnitude for the displacement for and $\eta \sim \gamma_1 \sim 0.2 Pas$, which are reasonable values for ferroelectric nematic liquid crystals[11]. We note that the direction of the movement of the defect in different domains can be opposite because the direction of the $curl\, \vec{P}$ can be opposite.

The elongation and separation of the defects along increasing electric fields indicate that they have opposite bound charges $\rho = -\nabla\vec{P}$ towards the opposite substrates. They are forced to move to the opposite direction along the external field as we observe experimentally (see Figure 6a,b). The separation of the bond charges involves the formation of a defect wall, as discussed by Lavrentovich[4]. We find that the charges split at a threshold field $E_s \sim 400 V/m$ when their separation is $s \sim 30\mu m$ (see Figure 6). From the balance of the electrostatic forces $qE_s = K\frac{q^2}{s^2}$, where $K = 9 \cdot 10^9 \frac{Nm^2}{C^2}$ , we get that $q \cong 4 \cdot 10^{-17}\,C$. Such value is much smaller than what we could expect from the divergence of the $|q| = \int \nabla P \cdot dV_d \sim \frac{2P}{w} w A_d = 2PA_d$, where $w$ is the thickness of the droplet and $A_d$ is the area of the defect. Assuming a defect area $A_d \sim 10 - 100\ \mu m^2$ and $P_o \sim 5 \cdot 10^{-2} C/m^2$, we expect $q \sim 10^{-12} - 10^{-11}\,C$. This indicates that most of the bound charges are screened out by free ions in the ferroelectric nematic fluid.

Since $E \cdot q \cong 10^3 \cdot 4 \cdot 10^{-17} \ll \gamma_{I-N_F} \cdot R \sim 3 \cdot 10^{-3} \cdot 3 \cdot 10^{-5}$, the deformation of the droplet shape is not due to the separation of the charges but is the result of the elastic oscillation of the deformed droplets. The deformation of the droplet shape requires that the ferroelectric energy density $P_O E$ be comparable to $\gamma_{I-N_F}/2R$, where $\gamma_{I-N_F}$ is the interfacial tension which typically an order of magnitude smaller than the surface tension at air[22] $\gamma_a \sim 3 \cdot 10^{-2} J/m^2$, and $2R$ is the diameter of the droplet. Indeed, with $P_o \sim 5 \cdot 10^{-2} C/m^2$ and $E \sim 10^{-3} V/m$ $P_o E \sim 50$, which overcomes $\gamma_{I-N_F}/2R$ for $R \geq 25\ \mu m$. This agrees with our observations where we see deformation of the R$\geq$ 30 $\mu m$ droplets only for $E < 1 mV/\mu m$ (see Figure 6).

What remains to be explained is the movement of the droplets along the electric field. As we see on Figure 5b, the speed is independent of the size of the droplets, proportional to the electric field and decreases at increasing frequencies (see inset to Figure 5b). These properties are typical of electrophoresis of colloid particles in isotropic electrolytes[23,24]. For particles (like our droplets) where $\frac{R}{\lambda_D} \gg 1$ the resulting electrophoretic velocity is given by the Helmholtz-Smoluchowski equation[23,24,25] $\vec{v}_{EP} = \frac{\varepsilon \cdot \varepsilon_o \zeta}{\eta_{iso}} \vec{E}$, where $\varepsilon_o = 8.854 \cdot 10^{-12} \frac{C^2}{J \cdot m}$ is the permittivity of the vacuum and $\zeta$ is the Zeta potential, which is equal to the electric potential at the slip surface near the solid (in our case the $N_F$ droplet) surface. According to our experimental results shown in the inset of Figure 5b the velocity decreases from about 10 μm/s at DC to about 3 μm/s at 2Hz. For typical zeta potential of $\zeta{\sim}100mV$ and $\eta_{iso}{\sim}0.1Pas$ and for $E = 10^3 V/m$ these electrophoretic values require that the dielectric constant of the isotropic fluid be $300 < \varepsilon < 1000$. Such values were indeed observed in the isotropic melt of several ferroelectric nematic materials[8]. The observed frequency dependence can be explained by the relaxation frequency $f_c$ of the screening free charges around the $N_F$ droplets which for $\frac{w}{\lambda_D} \gg 1$ can be given as $f_c = \frac{D}{w^2} {\sim} 1\ Hz$[26], where $D{\sim}2 \cdot 10^{-11} \frac{m^2}{s}$ is the diffusion constant in the isotropic liquid and $w{\sim}4\ \mu m$ is the width of the droplets.

To summarize, we have studied ferroelectric nematic liquid crystal droplets coexisting with the isotropic melt. We have found that the droplets have flat pancake-like shapes that are floating in the isotropic melt. In the center of the droplets a defect with low birefringence wing shape is present that moves perpendicular to a weak in-plane electric field, then it extends and splits in two at higher fields. In parallel to the defect motion and extension, the entire droplet drifts along the electric field with a speed that is independent of the size of the droplet and is proportional to the amplitude of the electric field. Above 1V/mm fields the shape of the droplets get deformed and oscillates between prolate and oblate ellipsoids due to the viscoelastic nature of the $N_F$ droplets. These observations have led us to determine the polarization field and revealed the presence of a pair of positive and negative bound electric charge due to divergences of polarization around the defect volume.

## 5. Acknowledgment

This work was supported by NSF DMR 2210083.

## Supporting Information

# Ferroelectric Nematic Droplets in their Isotropic Melt

*Kelum Perera[1,2], Rony Saha[1,2], Pawan Nepal[4], Rohan Dharmarathna[2], Md Sakhawat Hossain[2,3], Md Mostafa[2,3], Alex Adaka[2,3], Ronan Waroquet[2], Robert J. Twieg[4] and Antal Jákli[1,2,3]*

[1]Department of Physics, Kent State University, Kent OH, 44242, USA
[2]Advanced Materials and Liquid Crystal Institute, Kent State University, Kent OH, 44242, USA
[3]Materials Science Graduate Program, Kent State University, Kent OH, 44242, USA
[4]Department of Chemistry and Biochemistry, Kent State University, Kent, OH 44242, USA

The scheme used for preparation of the RT11064 used in this study is provided in Figure S1. Commercial 3,4-dimethoxybenzoic acid is converted to its acid chloride using oxalyl chloride in dichloromethane (a). This acid chloride is reacted with 2-fluoro-4-hydroxybenzaldehyde with pyridine as base in dichloromethane solvent (b). This two-ring aldehyde is oxidized to the analogous carboxylic acid using Oxone in DMF (c). This two-ring carboxylic acid is converted to its respective acid chloride using oxalyl chloride in dichloromethane (d). Finally, this acid chloride is esterified with 2-fluoro-4-hydroxynitrobenzene with pyridine as base and dichloromethane as solvent to give the desired three-ring product RT11064 (e). The first three steps in this scheme (a-c) have already been described in detail and are not reproduced here. [12]. The two-ring acid chloride obtained in the fourth step (e) is a new compound, but the method of preparation employed is routine. The final fifth step between this acid chloride and 2-fluoro-4-hydroxynitrobenzene with pyridine as base and dichloromethane as solvent gives RT11064. The yield in this final step of this specific prep was low but later preps of this same substance using a more careful chromatography provided a better yield. The RT11064 prepared by this route has spectroscopic properties very similar to the reported material prepared by an alternative route[3]. The final step of the synthesis was also attempted under Steglich and Mitsunobu conditions with low or no yield respectively. Details about the idiosyncrasies of this ester forming reaction will be described in more detail elsewhere.

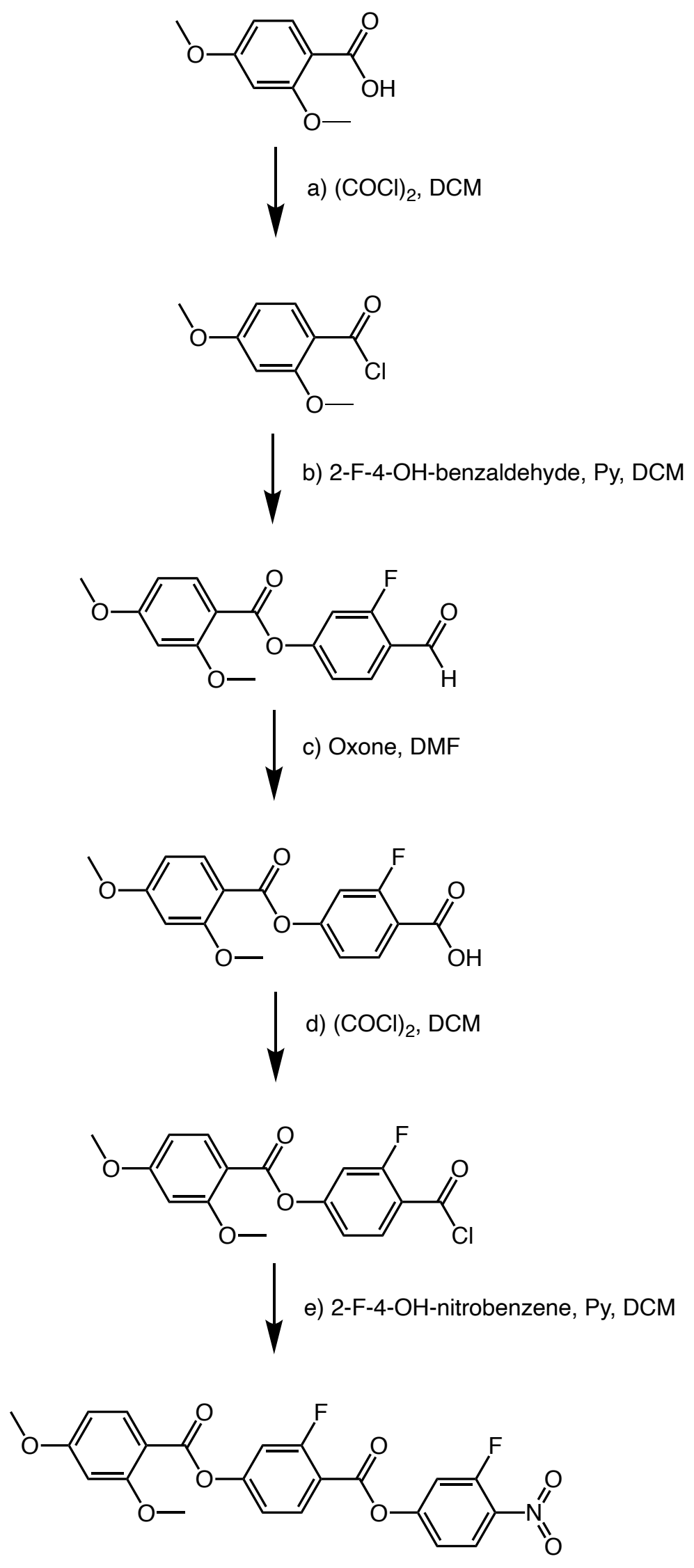

*Figure S1. Five Step Synthesis Scheme Utilized for the Preparation of RT11064 Benzoic acid, 2,4-dimethoxy-, 3-fluoro-4-[(3-fluoro-4-nitrophenoxy)carbonyl]phenyl ester*

Step (e)

3-Fluoro-4-(chlorocarbonyl)phenyl 2,4-dimethoxybenzoate

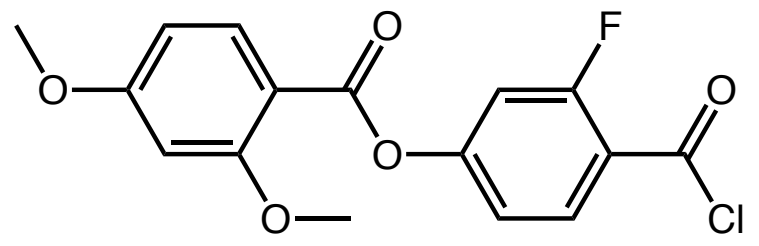

In a 250 ml round bottom flask with fitted with stirbar and bubbler was placed the 3-fluoro-4-carboxyphenyl 2,4-dimethoxybenzoate (1.60 gm, 5.0 mmol) and dry dichloromethane (40 ml). The slurry was stirred in a cold water dimethylformamide. Gas evolution ensued and the mixture cleared and was allowed to warm and was stirred overnight. The next day the dichloromethane and excess acid chloride were removed by rotary evaporation. Dry dichloromethane (20 ml) was added, and the mixture was concentrated again. The acid chloride (1.75 gm, 100%) was used directly in the subsequent esterification.

Step (f) RT11064

Benzoic acid, 2,4-dimethoxy-, 3-fluoro-4-[(3-fluoro-4-nitrophenoxy)carbonyl]phenyl ester

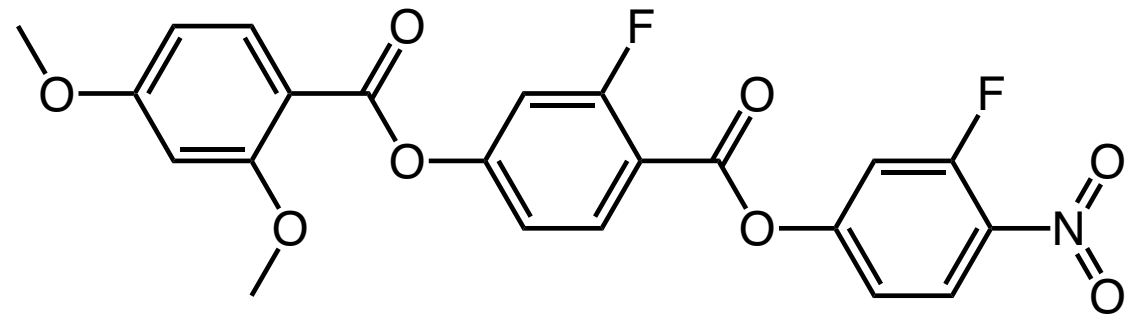

In a 500 ml recovery flask with stir bar was placed 3-fluoro-4-nitrophenol (1.17 gm, 7.50 mmol), and dry dichloromethane (40 ml). The mixture was stirred under nitrogen in an ice water bath and dry pyridine (2.40 gm, 30.0 mmol) was added all at once. The acid chloride (5.00 mmol) dissolved in dry dichloromethane (15 ml) was added dropwise and the resulting mixture was allowed to warm to room temperature and was stirred for 24 hr. The mixture was concentrated to about half its volume and applied to the top of a silica gel column (5 cm x 28 cm) made up with dichloromethane and eluted with dichloromethane. Analysis by TLC indicated that some intermediate fractions contained only pure product and they were combined, concentrated and the solid obtained was recrystallized from 1-propanol (0.591 gm, 26%). Additional product was present in the other chromatography fractions but was not further pursued.

$^{1}$H NMR (400 MHz, DMSO) δ 8.33 (t, *J* = 8.8 Hz, 1H), 8.21 (t, *J* = 8.5 Hz, 1H), 8.01 (d, *J* = 8.7 Hz, 1H), 7.78 (dd, *J* = 11.9, 2.4 Hz, 1H), 7.57 – 7.45 (m, 2H), 7.35 (ddd, *J* = 8.7, 2.3, 0.7 Hz, 1H), 6.78 – 6.65 (m, 2H), 3.89 (s, 6H).

$^{13}$C NMR (100 MHz, DMSO) δ 165.71, 163.89, 162.52, 162.23, 161.29, 160.76, 160.71, 157.00, 156.88, 155.69, 155.58, 154.38, 135.34, 135.27, 134.73, 133.94, 128.04, 119.74, 119.70, 119.48, 119.45, 114.36, 114.27, 113.45, 113.21, 112.41, 112.16, 109.66, 106.30, 99.47, 56.51, 56.24.

$^{19}$F NMR (376 MHz, DMSO) δ -105.98 (m, 1F), -115.26 (m, 1F).

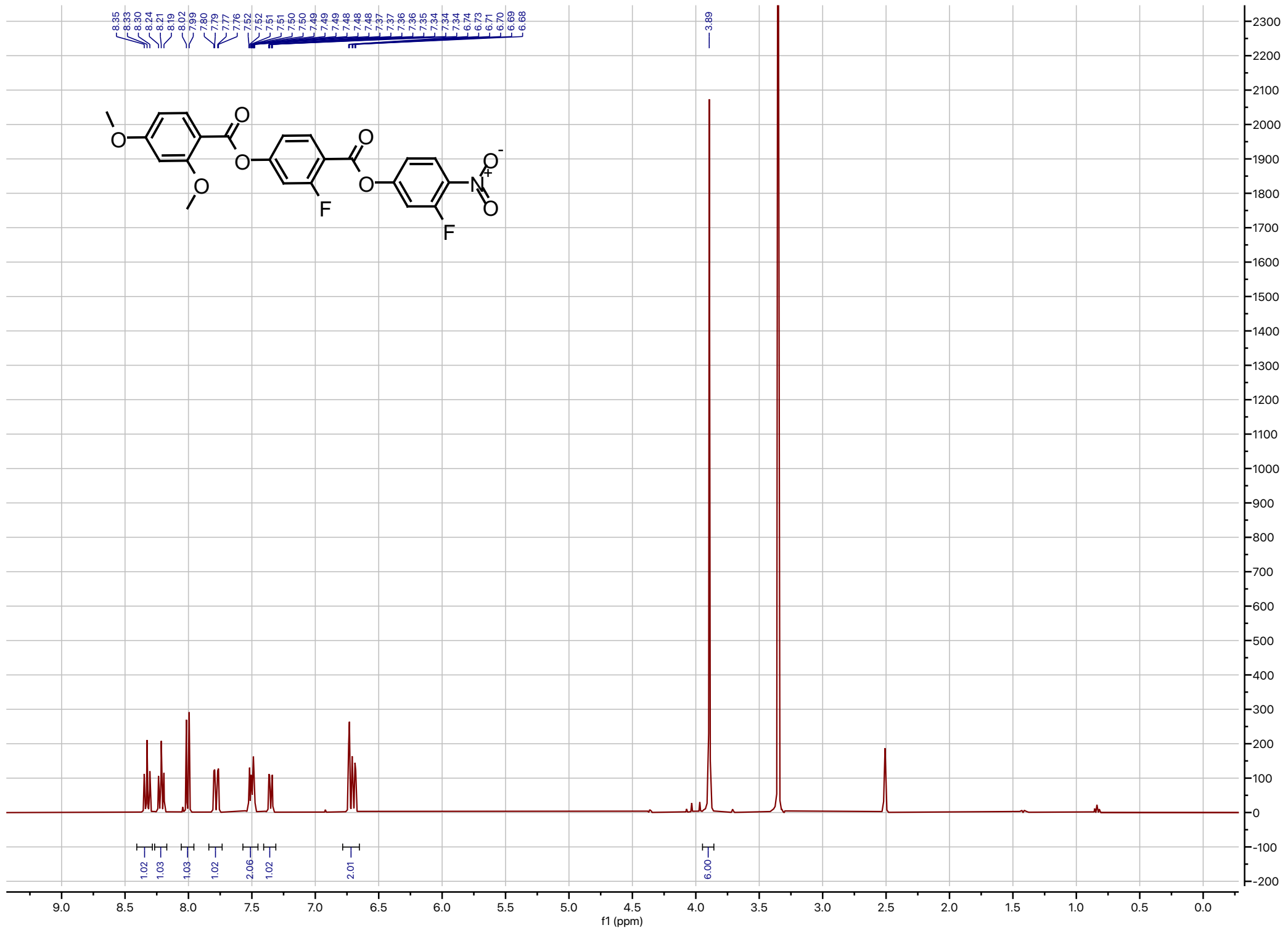

$^{1}$H NMR of RT11064

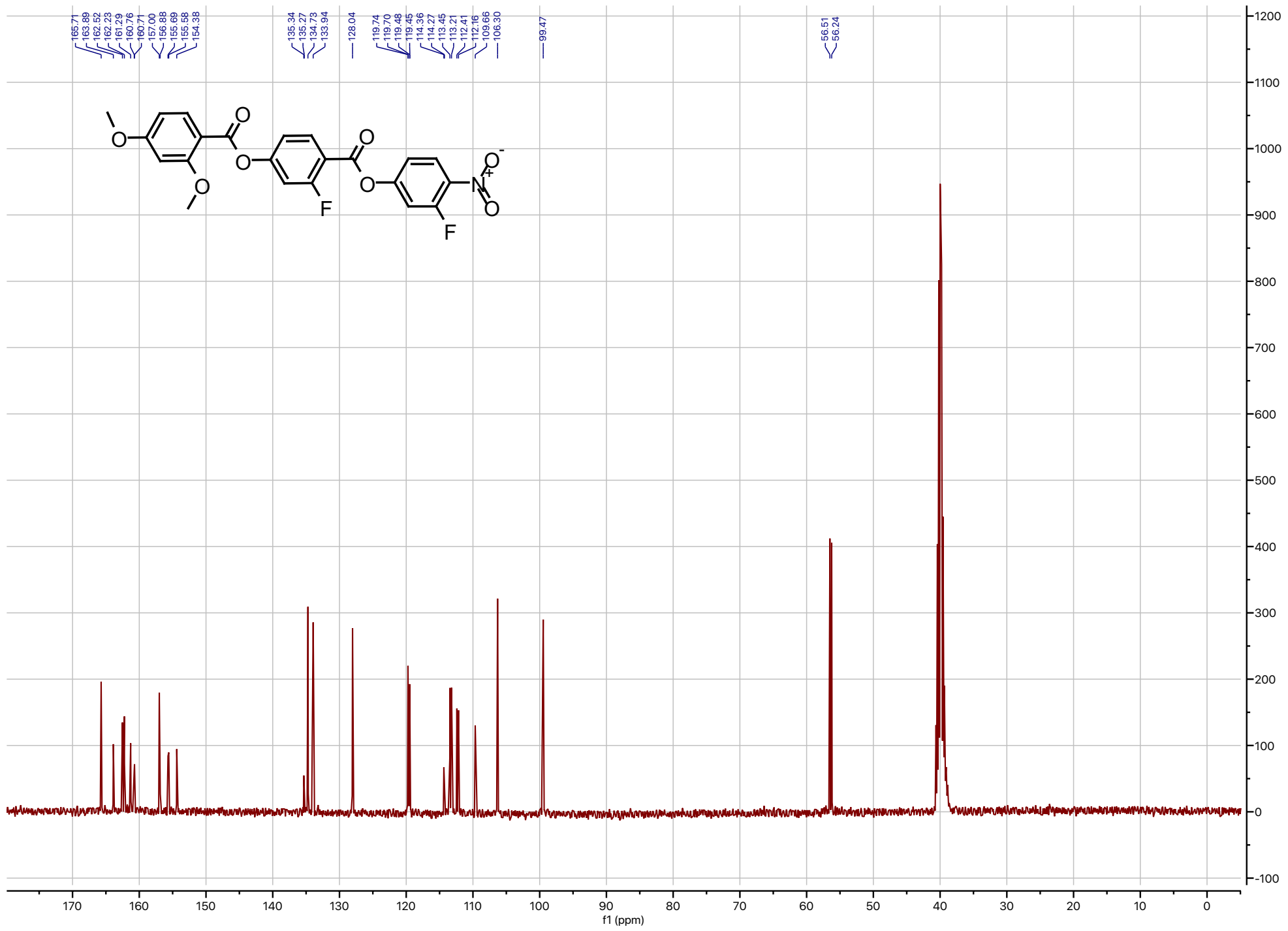

$^{13}$C NMR of RT11064

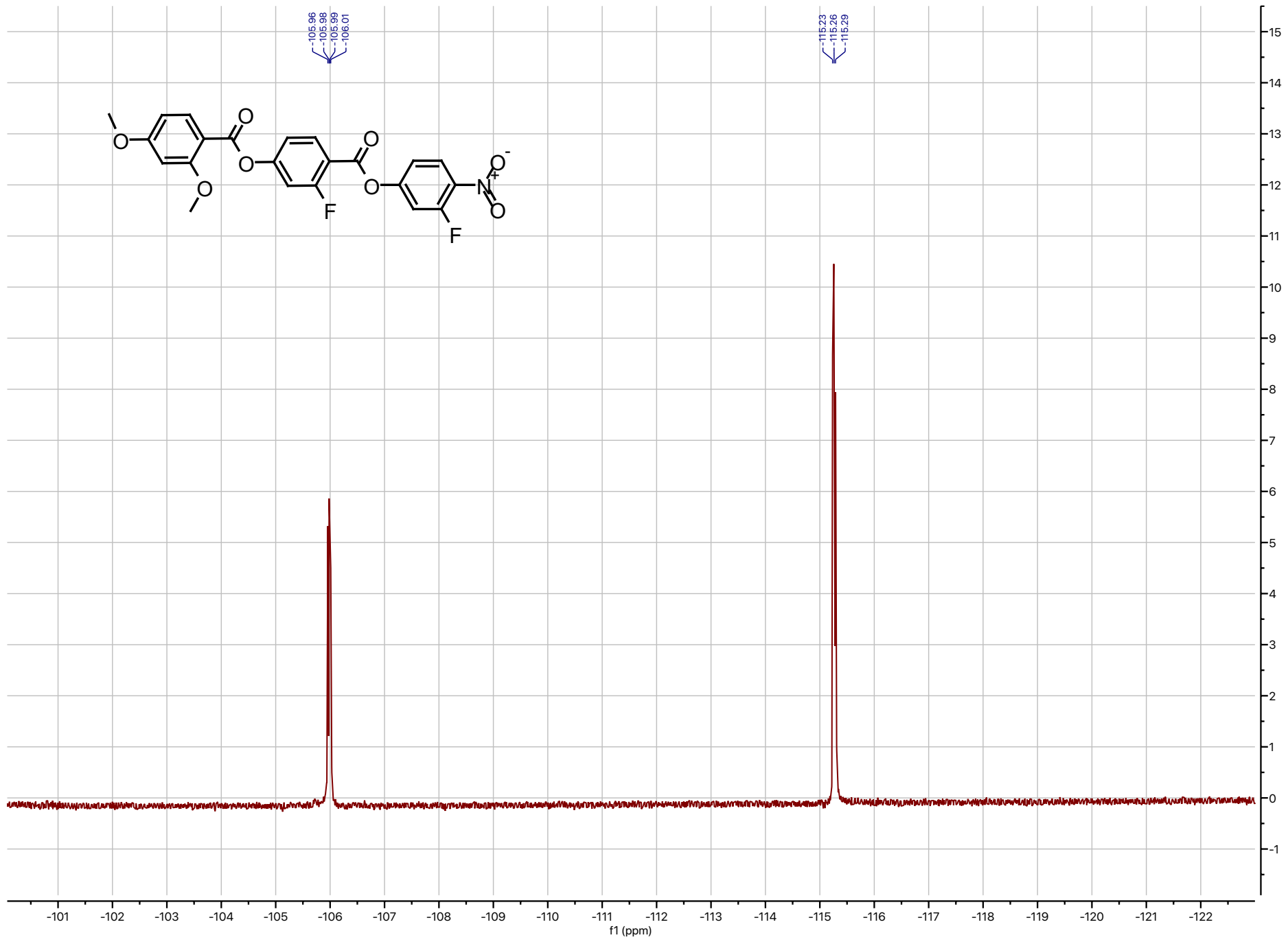

$^{19}$F NMR of RT11064

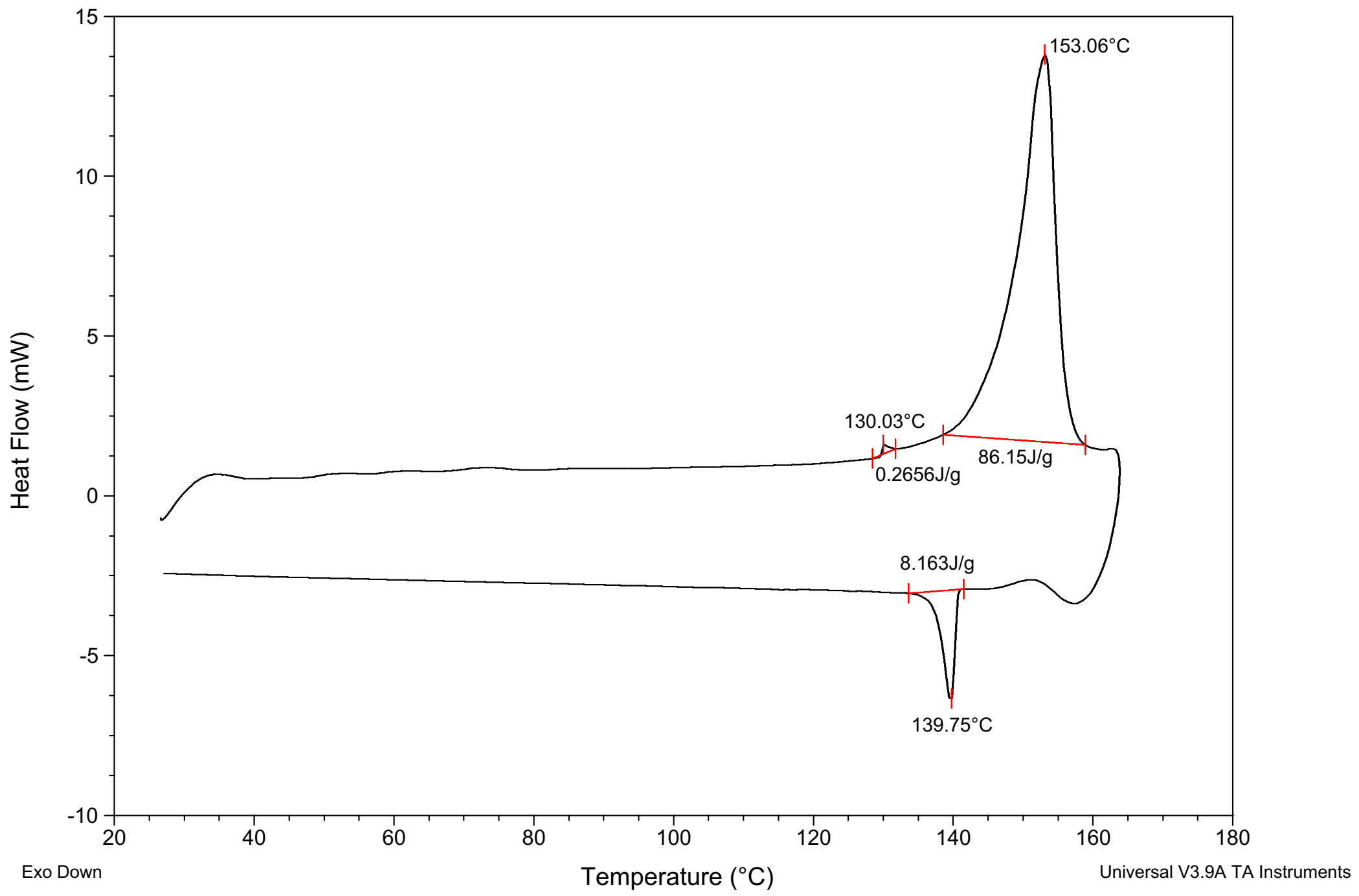

*Figure S2: DSC of RT11064. First heating and cooling cycle (heating and cooling rate = 5 ˚C/min. Sample size = 5.567 mg).*

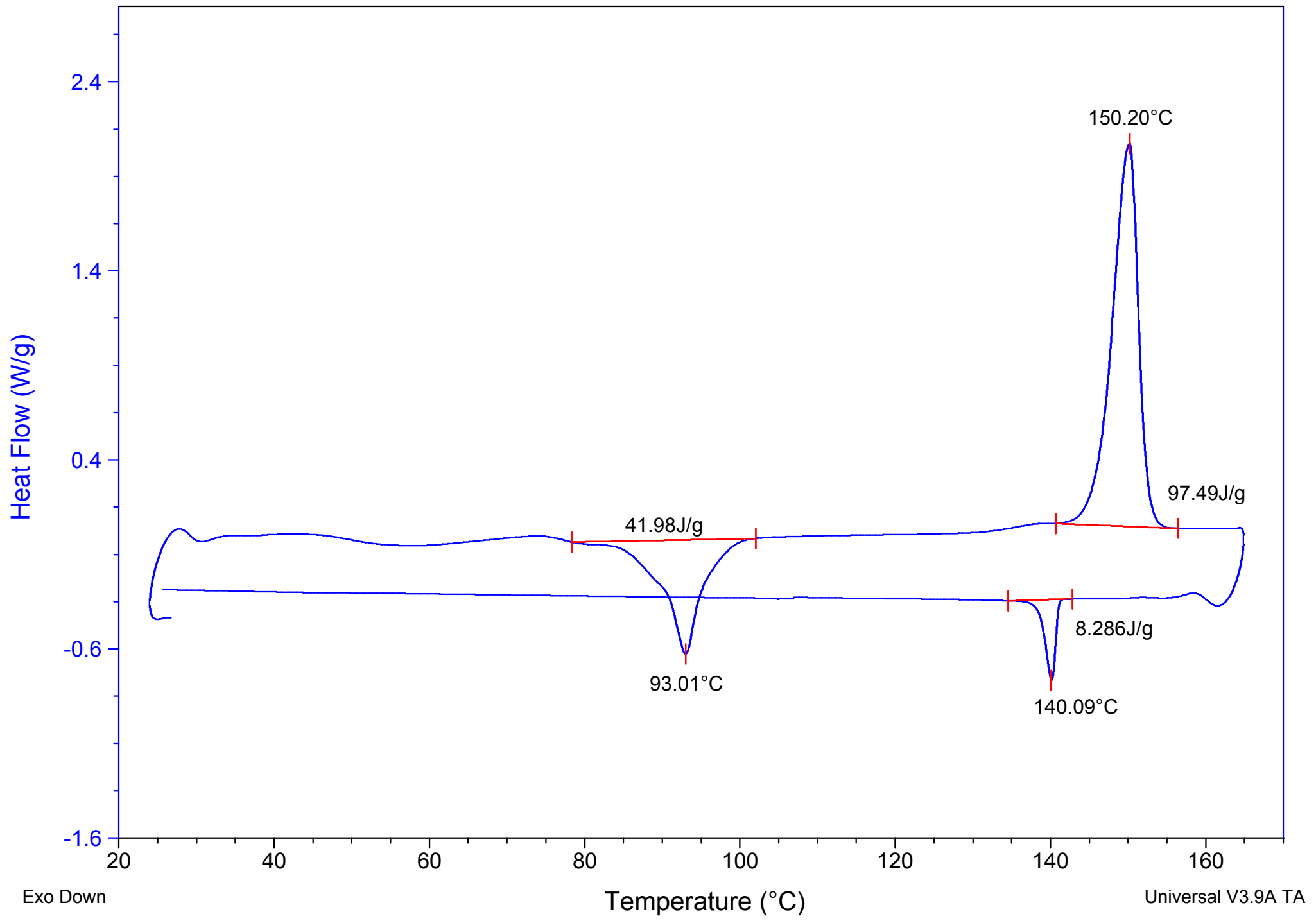

*Figure S3: DSC of RT11064. Second heating and cooling cycle (heating and cooling rate = 5 ˚C/min. Sample size = 5.567 mg).*

Video 1: Drift of the droplets along 0.5V DC electric voltage applied left (first part) and right (second part).

Video 2: Shape change of the droplets under 100 mHz, 1.5 V/mm electric field.

## SI References